\documentclass[letter,longauth]{aa}  
\usepackage{graphicx}
\usepackage{txfonts}
\usepackage[colorlinks=true,linkcolor=blue,citecolor=blue]{hyperref}
\usepackage[toc,page]{appendix}
\usepackage{float}
\usepackage{subfloat}
\usepackage{subcaption}
\usepackage[flushleft]{threeparttable}
\usepackage{overpic}
\usepackage{multirow}
\usepackage{cellspace} 
\usepackage[utf8]{inputenc}

\begin{document}

   \title{JWST reveals a supernova following a gamma-ray burst at z $\simeq$ 7.3}

   \subtitle{}

 \author{A.~J.~Levan\inst{1,2}\fnmsep\thanks{E-mail: a.levan@astro.ru.nl},
   B.~Schneider\inst{3},
   E.~Le Floc'h\inst{4},
   G.~Brammer\inst{5,6},
   N.~R.~Tanvir\inst{7},
   D.~B.~Malesani\inst{5,6},
   A.~Martin-Carrillo\inst{8},
   A.~Rossi\inst{9},
   A.~Saccardi\inst{4},
   A.~Sneppen\inst{5,6},
   S.~D.~Vergani\inst{10,11,12},
   J.~An\inst{13},
   J.-L.~Atteia\inst{14},
   F.~E.~Bauer\inst{15},
   V.~Buat \inst{3},
   S.~Campana \inst{12},
   A.~Chrimes\inst{16},
   B.~Cordier \inst{17},
   L.~Cotter\inst{8}, 
   F.~Daigne\inst{18},
   V.~D'Elia\inst{19},
   M.~De Pasquale\inst{20},
   A.~de~Ugarte Postigo\inst{3},
   G.~Corcoran\inst{8},
   R.~A.~J.~Eyles-Ferris\inst{7},
   H.~Fausey\inst{21},
   A.~S. Fruchter\inst{22},
   O.~Godet\inst{14},
   B.~P.~Gompertz\inst{23},
   D.~G\"otz\inst{4},
   N.~Habeeb\inst{7},
   D.~H.~Hartmann\inst{24},
   L.~Izzo\inst{25,26},
   P.~Jakobsson\inst{27},
   T.~Laskar \inst{28},
   A.~Melandri\inst{29},
   P.~T.~O'Brien\inst{7},
   J.~T.~Palmerio\inst{4},
   L.~Piro \inst{30},
   G.~Pugliese \inst{31},
   Y.~L.~Qiu\inst{13},
   B.~C.~Rayson \inst{7},
   R.~Salvaterra\inst{32},
   S. Schanne\inst{14},
   A.~L.~Thakur\inst{30},
   C.~C.~Th\"one\inst{33},
   D.~Watson\inst{5,6},
   J.~Y.~Wei\inst{13},
   K.~Wiersema\inst{34},
   R.~A.~M.~J.~Wijers\inst{31},
   L.~P.~Xin\inst{13},
   D.~Xu\inst{13},
   S.~N.~Zhang\inst{35}
   }

\institute{ %1
Department of Astrophysics/IMAPP, Radboud University Nijmegen, P.O.~Box 9010, Nijmegen, 6500~GL, The Netherlands
\and %2
Department of Physics, University of Warwick, Coventry, CV4 7AL, UK
\and %3
Aix Marseille Univ., CNRS, CNES, LAM, Marseille, France 
\and %4
Universit\'e Paris-Saclay, Universit\'e Paris Cit\'e, CEA, CNRS, AIM, 91191, Gif-sur-Yvette, France   
\and %5
Niels Bohr Institute, University of Copenhagen, Jagtvej 155, 2200, Copenhagen N, Denmark
\and %6
The Cosmic Dawn Centre (DAWN), Denmark
\and %7
School of Physics and Astronomy, University of Leicester, University Road, Leicester, LE1 7RH, UK
\and %8
School of Physics and Centre for Space Research, University College Dublin, Belfield, Dublin 4, Ireland
\and %9
Osservatorio di Astrofisica e Scienza dello Spazio, INAF, Via Piero Gobetti 93/3, Bologna, 40129, Italy
\and %10
LUX, Observatoire de Paris, Universit\'e PSL, CNRS, Sorbonne Universit\'e, Meudon, 92190, France
\and %11
Institut d’Astrophysique
de Paris, CNRS, UMR 7095, 98 bis bd Arago, F-75014 Paris, France
\and %12
INAF–Osservatorio Astronomico di Brera, Via E. Bianchi 46, 23807
Merate (LC), Italy 
\and
National Astronomical Observatories, Chinese Academy of Sciences, Beijing 100101, China
\and IRAP, Universit\'e de Toulouse, CNRS, CNES, Toulouse, France
\and Instituto de Alta Investigacion, Universidad de Tarapaca, Casilla 7D, Arica, Chile
\and
European Space Agency (ESA), European Space Research and Technology Centre (ESTEC), Keplerlaan 1, 2201 AZ Noordwijk, The Netherlands
\and CEA Paris-Saclay, Irfu/D\'epartement d’Astrophysique, 9111 Gif sur Yvette, France
\and Sorbonne Universit\'e, CNRS, UMR 7095, Institut d'Astrophysique de Paris, 98 bis bd Arago, F-75014 Paris, France
\and
Space Science Data Center (SSDC) - Agenzia Spaziale Italiana (ASI), Via del Politecnico snc, I-00133 Roma, Italy
\and %18
MIFT Department, University of Messina, 
Via F. S. D'Alcontres 31, Messina, Italy
\and % 19
Department of Physics and Astronomy,
Baylor University,
One Bear Place \#97316,
Waco, TX, 76798, USA
\and %20
Space Telescope Science Institute, 3700 San Martin Drive, Baltimore, MD21218, USA
\and %22
School of Physics and Astronomy and Institute for Gravitational Wave Astronomy, University of Birmingham, Birmingham B15 2TT, UK
\and %13
Clemson University, Department of Physics \& Astronony, Clemson, SC 29634, USA
\and %17
INAF, Osservatorio Astronomico di Capodimonte, Salita Moiariello 16, I-80121 Naples, Italy
\and %18
DARK, Niels Bohr Institute, University of Copenhagen, Jagtvej 155A, 2200 Copenhagen, Denmark
\and %20
Centre for Astrophysics and Cosmology, Science Institute, University of Iceland, Dunhagi 5, 107 Reykjavik, Iceland
\and %24
Department of Physics \& Astronomy, University of Utah, Salt Lake City, UT 84112, USA
\and %15 
INAF – Osservatorio Astronomico di Roma, Via Frascati 33, 00078 Monte Porzio Catone, (RM), Italy
\and
INAF -- Istituto di Astrofisica e Planetologia Spaziali, via Fosso del Cavaliere 100, I-00133 Rome, Italy
\and %16
Anton Pannekoek Institute of Astronomy, University of Amsterdam, P.O. Box 94249, 1090 GE Amsterdam, The Netherlands
\and %31
INAF—Istituto di Astrofisica Spaziale e Fisica Cosmica di Milano, Via A. Corti 12, 20133 Milano, Italy
%- ORCID: 0000-0002-9393-8078
\and %32
E. Kharadze Georgian National Astrophysical Observatory, Mt. Kanobili, Abastumani 0301, Adigeni, Georgia
\and %23
Centre for Astrophysics Research, University of Hertfordshire, Hatfield, AL10 9AB, UK
\and
Key Laboratory of Particle Astrophysics, Institute of High Energy Physics, Chinese Academy of Sciences, Beijing 100049, China
}

   \date{Received September XX, YYYY; accepted March XX, YYYYY}

% \abstract{}{}{}{}{} 
% 5 {} token are mandatory
 
  \abstract
  {The majority of energetic long-duration gamma-ray bursts (GRBs) are thought to arise from the collapse of massive stars, making them powerful tracers of star formation across cosmic time. Evidence for this origin comes from the presence of supernovae in the aftermath of the GRB event, whose properties in turn link back to those of the collapsing star. In principle, with GRBs we can study the properties of individual stars in the distant universe. Here, we present \textit{JWST}/NIRCAM observations that detect both the host galaxy and likely supernova in the \textit{SVOM} GRB\,250314A with a spectroscopically measured redshift of $z\simeq 7.3$, deep in the era of reionisation. The data are well described by a combination of faint blue host, similar to many $z \sim 7$ galaxies, with a supernova of similar luminosity to the proto-type GRB supernova, SN~1998bw. Although larger galaxy contributions cannot be robustly excluded, given the evidence from the blue afterglow colours of low dust extinction, supernovae much brighter than SN~1998bw can be.
These observations suggest that, despite disparate physical conditions, the star that created GRB 250314A was similar to GRB progenitors in the local universe. }
   
   
   
   


   \titlerunning{JWST SN in GRB\,250314A}
   \authorrunning{Levan et al. }

   \maketitle
%
%-------------------------------------------------------------------

\section{Introduction}

Understanding stars in the early universe remains one of the prime goals in contemporary cosmology. The first generations of massive stars 
are the likely engines that drove cosmic reionization beyond $z \sim 6$ \citep[e.g.][]{reion} and were responsible for early cosmic chemical enrichment \citep[e.g.][]{kobayashi17}. The very different physical conditions at early cosmic epochs could plausibly result in markedly different stellar and binary evolution and hence varied stellar properties at the end of their lives \citep[e.g.][]{fryer19}. Enormous effort has been invested in studying the integrated light from populations of these stars, with progress from {\em JWST} particularly impressive in recent years \citep[e.g.][]{bunker23,curti23,robertson23,carniani24,witstok25,naidu25}. However, there are few probes of individual stars at this epoch because of the extreme luminosity distances involved. Observing massive stars as they collapse is one route to studying individual objects at these redshifts. Indeed, {\em JWST} surveys are now identifying supernovae in repeatedly observed
deep fields in increasing numbers \citep{decoursey25}, and likely out to 
$z \sim 5$ \citep{siebert24,decoursey25b}. 
Long gamma-ray bursts also offer a very promising avenue to study massive stars in the early universe. 
They are bright enough to be detected at very high redshifts $z>6$ \citep{haislip06,tanvir09,salvaterra09,Tanvir18,rossi22b}, theoretically up to $z\sim 20$ \citep{Lamb2000} with current technology, and offer bright backlights, enabling us to measure redshifts and dissect the properties of the interstellar medium in the host galaxy in exquisite detail \citep[e.g.][]{tanvir19,heintz23,saccardi23,saccardi25}. 

We can use GRB-supernovae to study their progenitors and, hence, individual stars at high redshift because of the connection between (most) long-duration GRBs and massive stars. Long GRBs are generally seen in association with
highly energetic broad-lined type Ic supernovae  \citep[SNe; e.g.][]{galama98,hjorth03}. 
In a minority of cases no supernova is seen, perhaps indicating direct collapse events \citep[e.g.][]{fynbo06}, although more recently the detection of kilonova signatures accompanying some low-$z$ long-GRBs argues that these are likely 
compact object mergers \citep[e.g.][]{rastinejad22,troja22,levan24,yang24}. Nonetheless, we expect the typically luminous GRBs observed at high redshift to be dominated by core collapse events, given that collapsar GRBs reach higher luminosities, and have shorter average elapsed times from initial star formation to explosion. 

GRB supernovae to date have been observed to $z<1$, but it is notable that the supernovae appear similar even when the energetics of the prompt GRB vary by 7 orders of magnitude \citep{levan14,melandri14,blanchard23}.
For example, there is remarkably little spread in their peak luminosities, with peak $M_V = -19.2 \pm 0.4$ reasonably describing the bulk of the population \citep[e.g.][]{Cano17,deugartepostigo18,melandri14}, although in at least one case, of an ultra-long duration GRB, a more luminous supernova, comparable to the superluminous supernovae, was observed \citep{greiner15,kann19}.
A pressing question is then whether this relative lack of diversity reflects uniformity in the progenitors, and whether this similarity holds across cosmic time, where physical conditions may be very different. Here we address this with observations of a supernova accompanying GRB\,250314A, at a spectroscopic redshift of $z\simeq7.3$. 

\begin{figure*}[h!]
\centering
\includegraphics[angle=0,width=12cm]{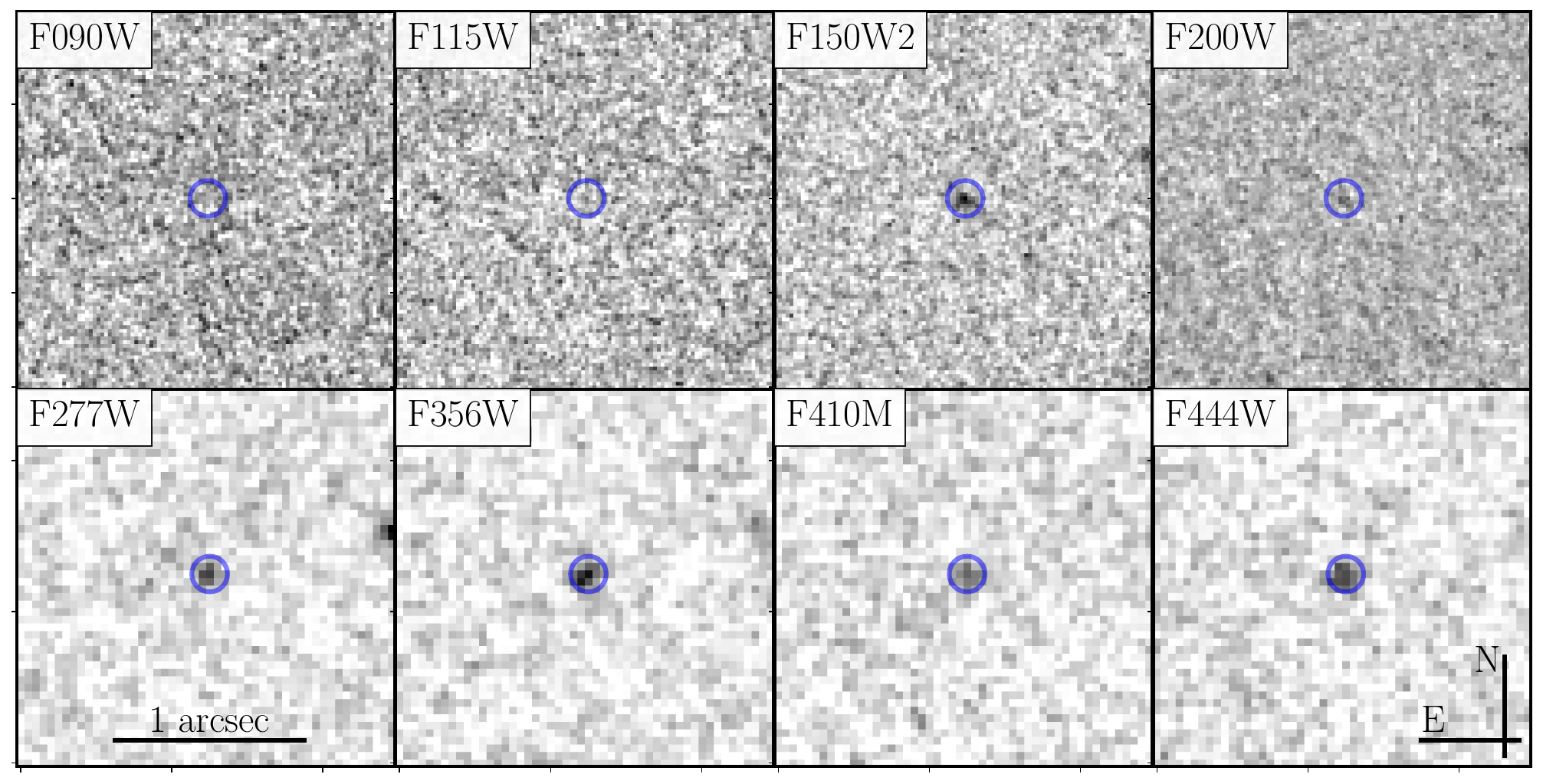}
\caption{Our 8-band NIRCAM images of the location of GRB\,250314A. The source is undetected in the bluest F090W band, consistent with the $z>7$ origin. A clear signature from an extended object is present in the F150W2 observations (with a very broad filter), with a blue colour between F150W2 and F200W. The source is much brighter in the NIRCAM red channel, and brightens consistently between F277W and F444W. 
}
\label{fig:mosaic}
\end{figure*}

\section{Observations}

\subsection{Burst discovery and redshift measurement} 
GRB\,250314A was detected by the ECLAIRs  instrument on the Space Variable Objects Monitor (\textit{SVOM}) satellite at 12:56:42 UT on 14 March 2025 \citep{2025GCN.39719....1W} and was also seen by its Gamma-Ray Monitor  \citep[GRM;][]{2025GCN.39746....1S}. Observations with the Visible Telescope (VT) did not reveal any counterpart \citep{2025GCN.39728....1L}, but an infared afterglow was discovered with the Nordic Optical Telescope \citep{2025GCN.39727....1M}. An X-ray counterpart was also identified by both the {\em Swift} X-Ray Telescope (XRT) \citep{2025GCN.39734....1K} and the {\em Einstein Probe} Follow-up X-ray Telescope (FXT) \citep{2025GCN.39739....1T}. Optical spectroscopy obtained at the VLT revealed a redshift of $z\simeq7.3$ based on a strong spectral break between the optical and IR regimes \citep{2025GCN.39732....1M}, a result later confirmed via deep $z-$band imaging \citep{2025GCN.39743....1R}. A full summary of the multi-wavelength follow-up of this event is presented in a companion paper \citep{cordier25}. 

\vspace{-2ex}

\subsection{JWST NIRCAM observations}
We obtained observations with \textit{JWST}/NIRCAM on 1 July 2025, an epoch $\sim 110$ days post burst (or 13 days in the rest-frame).
The 8-band imaging of the field is shown in Figure~\ref{fig:mosaic}, and photometry is shown in table~\ref{NIRCAM_phot}. At the location of the counterpart (see appendix) we find a faint source visible in all filters redward of F150W2. The source shows a near flat spectrum from F150W2 to F277W, followed by a strong rise to the redder filters. In F150W2 there is some weak evidence for extension of the source, suggesting that at least some of the light at this wavelength is contributed by the host galaxy. 

\section{A benchmark expectation for a GRB-supernova}
The light observed from a stellar collapse driven GRB at any given time can be split into components from the GRB afterglow, the associated supernova and the underlying host galaxy. On both theoretical grounds, and based on the extensive observations of GRBs to date we can make a prediction about a benchmark model for GRB\,250314A, adopting the assumption that it looks like GRBs locally -- this is our simplest model, and the rationale of our \textit{JWST} observations was to confront this model with observations. We describe these expectations for each of the components briefly below. 

There are relatively limited observations of the afterglow light in GRB\,250314A. However, the observations reported in \cite{cordier25} show that the afterglow is described as a power-law in time and frequency $F_{\nu} \propto t^{-\alpha} \nu^{-\beta}$ with $\alpha = 2.1 \pm 0.6$ and $\beta = 0.2 \pm 0.4$. This gives a large range of plausible magnitudes at the time of the \textit{JWST} observations, but all are faint $31 < F150W2 < 37$ mag. We therefore do not believe that there is a significant afterglow contribution in our data.

\begin{figure}
\centering
\includegraphics[angle=0,width=\columnwidth]{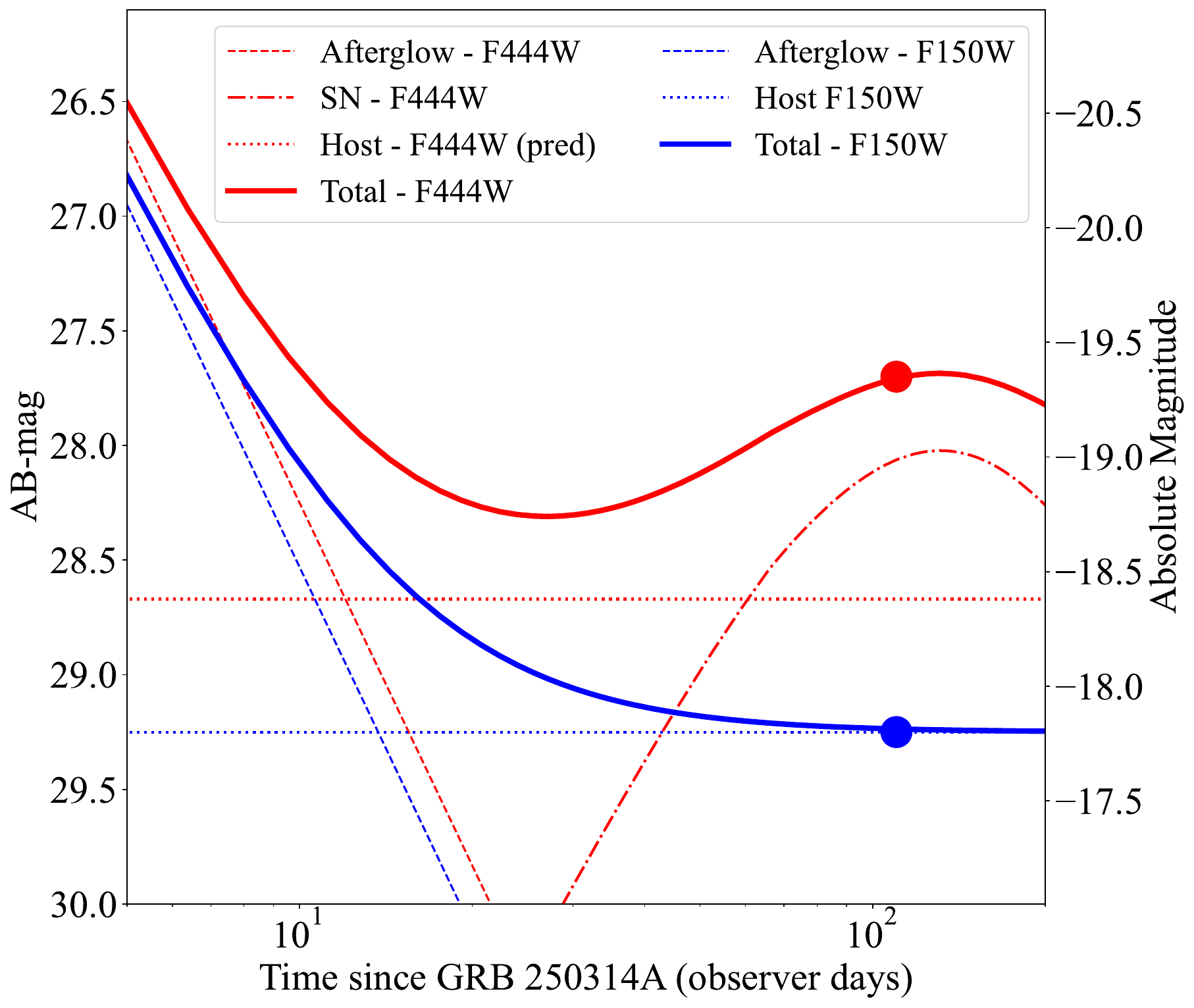}
\caption{The lightcurve of GRB\,250314A as expected in the F150W2 and F444W bands. Components from both afterglow, supernova and host galaxy are included, Expectations for the afterglow are based on the ground based $J$ and $H$ measurements, extrapolated to F444W based on the measured spectral slope. The host galaxy magnitudes are similarly based on the F150W2 detection (see text) and extrapolated to F444W based on the typical galaxy spectrum at $z \sim 7$. The observed points are also indicated, demonstrating luminosity consistency between observations and a SN~1998bw-like SN, in this case scaled to 70\% of the luminosity of SN~1998bw.}
\label{fig:lightcurve}
\end{figure}

\begin{figure}
\centering
\includegraphics[angle=0,width=\columnwidth]
{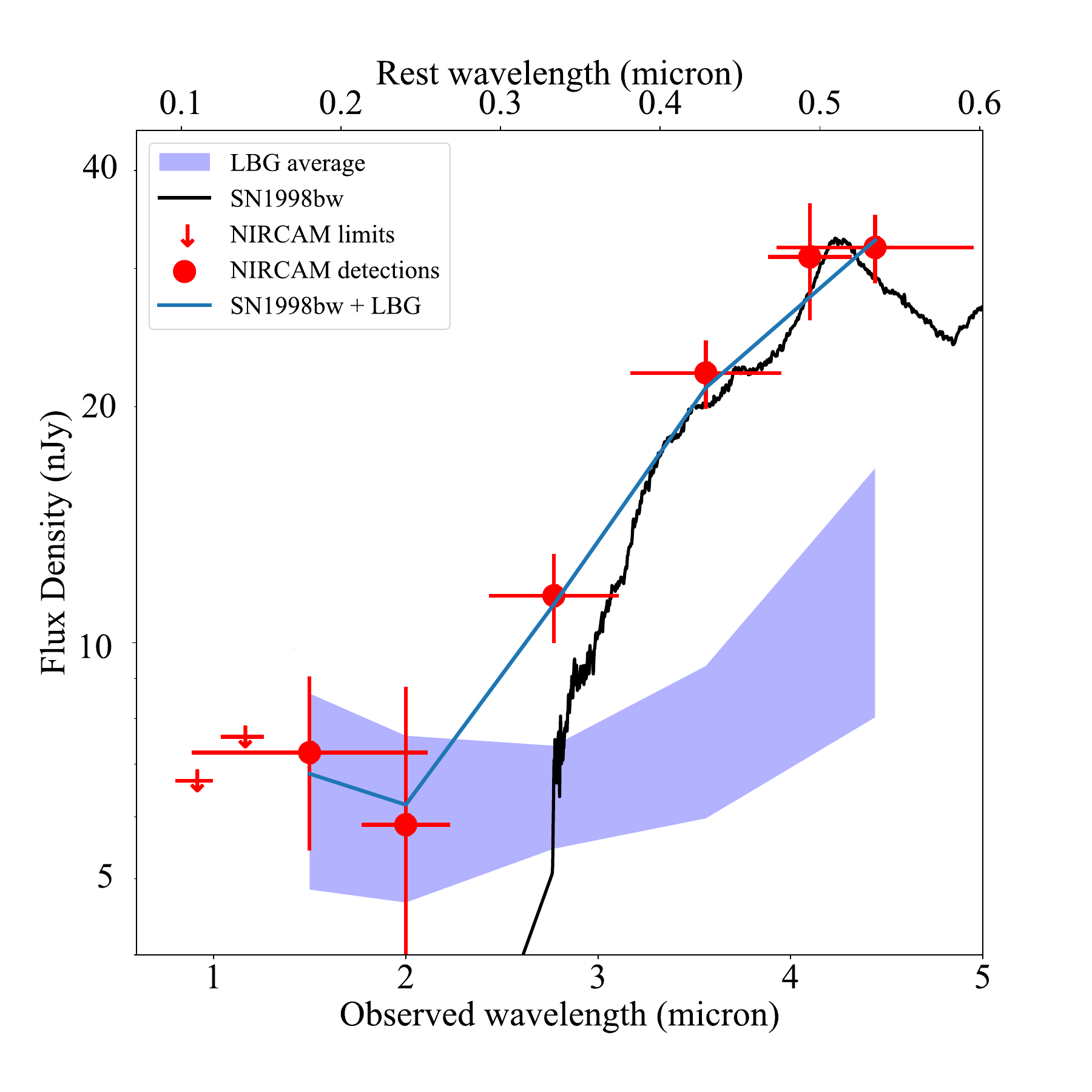}
\caption{The observed spectral energy distribution (SED) of the location of GRB\,250314A as observed with NIRCAM at 110 days post burst, roughly 13 days rest-frame (red points). The weak detection in the blue of a marginally extended source is consistent with a faint host galaxy, while the rise to the red is entirely consistent with the expectations of SN~1998bw at this epoch (black line). For comparison we also show (shaded region) the typical range of SEDs of the Lyman-break galaxies (LBGs) at $z \sim 7$ from \cite{Merlin2024}, and an SED obtained from summing an average LBG and a supernova with 70\% of the luminosity of SN\,1998bw (blue line).} 

\label{fig:sed}
\end{figure}

GRB supernovae are a relatively homogeneous population, showing peak magnitudes that typically only span a factor of two \citep{Cano17}. Indeed, SN~1998bw, the prototype of GRB supernovae is generally a good description of the SN seen in more distant events, and we adopt this here. Critically, SN Ic-BL exhibit heavy metal line blanketing blueward of $\sim 3000$\,\AA\,in the rest-frame. Hence supernova light should be undetected in the blue NIRCAM filters, but should dominate in the red, with SN~1998bw \citep{galama98} reaching a peak magnitude of $M_B \sim -19.3$ (F444W=27.7(AB)) at $z=7.3$. These expectations are shown, along with our observations, in Figures~\ref{fig:lightcurve} and \ref{fig:sed}.

The host galaxy is the most difficult element to constrain. Observations of local GRB hosts show them to be low luminosity, relatively compact and highly star-forming galaxies \citep{fruchter06,savaglio09,perley16,schneider22}. At high redshift, the majority of hosts are undetected, even to the limits of {\em HST} \citep{tanvir12,sears24}. Our expectation is hence of a faint, blue galaxy, probably with colour similar to typical Lyman-break galaxies at $z \sim 7$ \citep{roberts24}, but whose luminosity could span a range from undetectable to dominating the observed light \citep[see also][]{salvaterra13}. The range of GRB host galaxy magnitudes from several large scale space-based surveys are shown in Figure~\ref{fig:host}.

\section{The supernova in GRB\,250314A}
 Overall, the comparison between NIRCAM photometry and the canonical model is extremely good. We detect a marginally extended object in F150W2, suggesting a faint host galaxy with $M_{\rm UV}\approx-17.8$. This is somewhat fainter than the handful of high-$z$ GRB hosts that have been detected to date \citep{mcguire16,rossi22}, but is consistent with the upper limits that have been placed on others \citep{tanvir12}. The source then exhibits a strong rise through the F277W to F444W bands, reaching a peak at
$\mathrm{F444W}=27.64$, corresponding to an absolute magnitude of $M_{\rm B} =-19.41$ mag. The observing epoch of 110 days is $\sim 13$ days in the rest-frame, approximately the time of the B-band peak of SN~1998bw (14 days, \citealt{galama98,Clocchiatti11}). 
If we adopt a typical Lyman break galaxy spectral shape from \cite{Merlin2024} and use the F150W2 observations to fix the normalization, the resulting absolute magnitude of probable transient emission (host subtracted) is $M_B = -18.9$, which is $\sim 70$\% of SN~1998bw at the same redshift.

Hence, the photometry is entirely consistent with the predictions made prior to the observations, and is supportive of the detection of  supernova light at $z \simeq 7.3$. 
This supernova has a similar luminosity and spectral shape to SN~1998bw, and is also consistent with the range of supernova properties seen in GRBs locally \citep[e.g.][]{Cano17,melandri14}. This is a reasonable interpretation of the observed data, although it should be noted that the spectral shape of SN\,1998bw is explained by metal line blanketing in the rest-frame UV, which is a feature in the spectra of many supernova sub-types. Hence, these observations do not strongly constrain the type of the associated supernova, but do rule out a much bluer or brighter event than SN\,1998bw.

There are expectations that supernovae at high redshift may appear different to those in the local universe, either because of the general changes to stellar (and binary) evolution with metallicity \citep[e.g.][]{yoon06,eldridge19}, because of the very different evolutionary channels that metal free (so-called pop III) stars may take \citep[e.g.][]{whalen13}, or indeed differences in the stellar population properties such as the initial mass function, feedback, or the availability of different binary evolution channels. At low metallicity stars retain both more mass and angular momentum, potentially GRBs with higher supernova ejecta masses, and hence brighter associated supernovae. While the most massive pop III stars likely explode as pair instability events, some are also of sufficiently low mass to leave compact objects, and may result in very long-lived supernovae, that, due to the lack of metal line blanketing, may be much bluer than supernovae seen locally \citep[e.g.][]{whalen13}. 

In the case of GRB\,250314A this would not appear to be the case. The similarity of the supernova to SN~1998bw implies that the progenitor of GRB\,250314A is similar to that of the GRBs that we observe in the local universe. In particular, these observations robustly constrain any more luminous events (formally the SN could be less luminous in the case of a larger host contribution, see below). Whether high-$z$ GRBs exhibit the same homogeneity in supernova behaviour as is seen locally will ultimately be something that can be demonstrated via the observations of larger samples. 
 
\section{Alternative interpretations}
Although the observations match the expectations of the canonical model, they do not, at present, demonstrate the variability of the source, and it is relevant to consider if they may also be described by alternative solutions. Perhaps the most pressing of these is if the light could be entirely dominated by the host galaxy with a smaller (or even zero) contribution from the underlying supernova. 

Further analysis of this scenario is given in the appendix, but in summary, the observed spectral energy distribution is not a good match to the SEDs of the large number of galaxies now found at $z \sim 7$ by \textit{JWST}  \citep[e.g.][]{Merlin2024} due to the very red colours in the NIRCAM bands. Moreover, it is too blue (in particular in the F200-F277W region of the spectrum) to be a
match to the population of ``little red dots" (LRDs) at similar redshifts. Given the limited number of data points and significant degrees of freedom, it is possible to obtain a reasonable fit via SED fitting with CIGALE \citep{Boquien2019}. However, these fits do require a substantial population of older stars that again places them apart from the galaxy population at $z \sim 7$. 

Critically, in the galaxy scenario, the similarity in luminosity of the red component to SN~1998bw must be coincidental. Given these considerations, we conclude that it is most likely that the NIRCAM observations are capturing the combination of a blue host and associated supernova.

\section{Conclusions}
We have presented the discovery of a likely supernova accompanying GRB\,250314A at $z \simeq 7.3$. 
These observations demonstrate the promise of \textit{JWST} to identify stellar explosions in the very early universe, especially relatively bright supernovae, such as those associated with GRBs. In the case of GRB\,250314A, our preferred model is one in which the associated supernova is indistinguishable from the supernovae seen in GRBs at $z<1$, implying little evolution in GRB-supernova properties across much of cosmic history. 
This conclusion is similar to studies showing that the properties of high-$z$ GRB afterglows are consistent with those at lower redshifts \citep{laskar14}.
Thus, it may well be that the mechanisms and stars that give rise to GRBs at $z \sim 0$ are the same as those at $z \sim 7$.

\begin{acknowledgements}
Acknowledgements can be found in the appendix.

\end{acknowledgements}

\vspace{-1cm}

\bibliographystyle{aa} % style aa.bst
\bibliography{refs.bib} % your references 

\begin{appendix}
\section{Photometry}
NIRCAM frames were retrieved from the \textit{JWST} archive, and subsequently re-drizzled to a pixel scale of 0.02 and 0.04 arcsec/pixel for the blue and red cameras, respectively. At $\sim$110~days after the GRB trigger (13 days in rest-frame), we obtained 1868\,s of observations in the F090W, F115W, F200W, 356W, F410M and F444W filters and 3800\,s of exposure with the F150W2 and F277W filters providing complete coverage from the 1 to 5 micron regime (roughly from Ly-$\alpha$ to 0.6 micron at a redshift of 7.3). 

Photometry was undertaken in 0.08 and 0.12 arcsecond apertures for the blue and red channel respectively, and subsequently corrected for encircled energy losses assuming a point-like source. The resulting photometric measurements are shown in Table~\ref{NIRCAM_phot}. 

\begin{table}[h!]     
\centering
\caption{Log of \textit{JWST} observations. Upper limits are given at the $2\sigma$ level and other uncertainties at 1$\sigma$} 
%\small
\begin{tabular}{c c c c}       
\hline
Date & $\Delta$ T (days) & Filter & Magnitude\\
\hline
2025-07-01 23:23:35 & 109.43  & F090W & $>29.34$ \\
2025-07-01 22:41:32 & 109.41 & F115W & $>29.20$ \\
2025-07-01 21:27:16 & 109.35 & F150W2 & $29.25 \pm 0.11$ \\
2025-07-02 00:05:49 & 109.46  & F200W & $29.48 \pm 0.50$ \\
2025-07-01 21:27:16 & 109.35  & F277W & 28.75 $\pm$ 0.14 \\
2025-07-01 22:41:32 & 109.41 & F356W  & 28.04 $\pm$ 0.09 \\
2025-07-01 23:23:35 & 109.43  & F410M & 27.67 $\pm$ 0.15 \\
2025-07-02 00:05:49 & 109.46 & F444W  & 27.64 $\pm$ 0.11 \\
\hline
\end{tabular}
\label{NIRCAM_phot} 
\end{table}

\section{Astrometry}
We locate the afterglow on the images by performing relative astrometry between the HAWK-I J-band image presented in \cite{cordier25}, and the NIRCAM observations. Using 10 objects in common between the NIRCAM F150W2 and HAWKI observations we obtain an astrometric match with an RMS of 1.8 NIRCAM pixels (0.037 arcseconds). We hence determine that the offset between the NIRCAM and ground-determined positions is $1.5 \pm 1.8$ pixels, confirming that the \textit{JWST} source is consistent with the afterglow. There is no significant or systermatic centroid shift in the location of the transient between the source location measured in the different NIRCAM bands, although the modest signal to noise of the detections makes robust statements in this regard challenging. 

\section{Host galaxy contribution and properties}

\begin{figure}
\centering
\includegraphics[angle=0,width=\columnwidth]{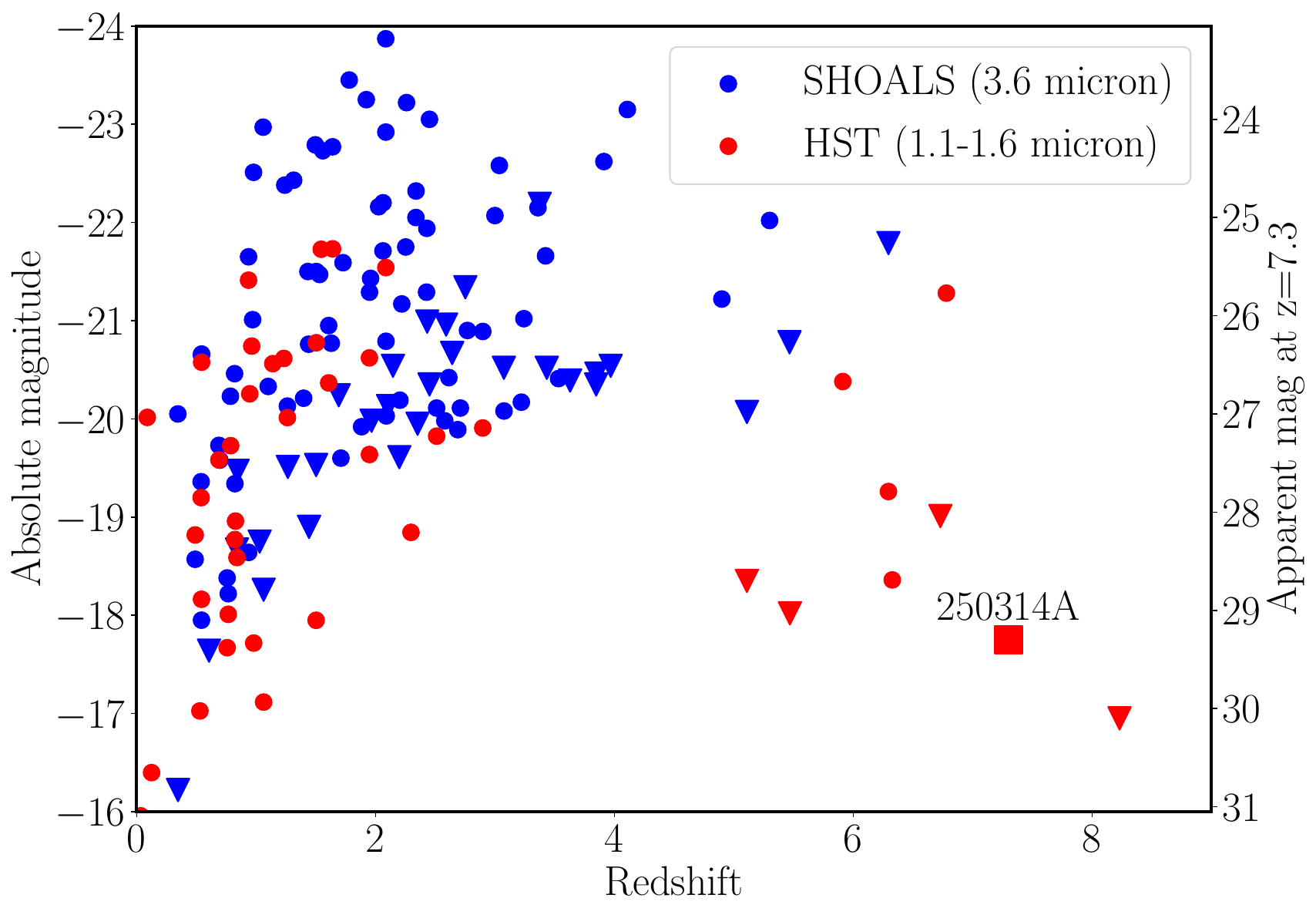}
\caption{Comparison of GRB host galaxy absolute magnitudes (or upper limits) from the SHOALS sample at 3.6 microns \citep{perley16}, and from various {\em HST} observations at 1.1-1.6 microns \citep{tanvir12,mcguire16,lyman17} (note that these are in a fixed observed band, and so substantially different rest-frame band at different redshifts). Under the assumption that the host of GRB 250314A dominates the emission in the F150W2 band, the observations are consistent with other GRB hosts at high-$z$. 
}
\label{fig:host}
\end{figure}

As noted in the main text, a key question regarding the emission is if the observed spectral energy distribution can be explained entirely by the presence of a host galaxy. It is apparent that the F150W2 observations can be explained as a host galaxy, as also evidenced by the apparent small extension of the source. The magnitude of this host galaxy, compared to observations of other high-$z$ hosts by {\em HST} is consistent with expectations (see Figure~\ref{fig:host}, although we note that F150W2 = 29.25 is beyond the limits of HST for all but the longest exposures). However, in our preferred interpretation, the redder observations, and the shape of the SED are explained by the presence of a supernova component.  

To assess if a host galaxy provides a viable alternative to a supernova we use two different approaches, the first is an direct empirical comparison to galaxies at $z \sim 7$ that have been detected in increasing numbers by \textit{JWST} in recent years. The second is to fit the SED with a population synthesis modelling code (in this case CIGALE) to ascertain which kinds of stellar populations could explain the emission. 

\begin{figure}
\centering
\includegraphics[angle=0,width=\columnwidth]{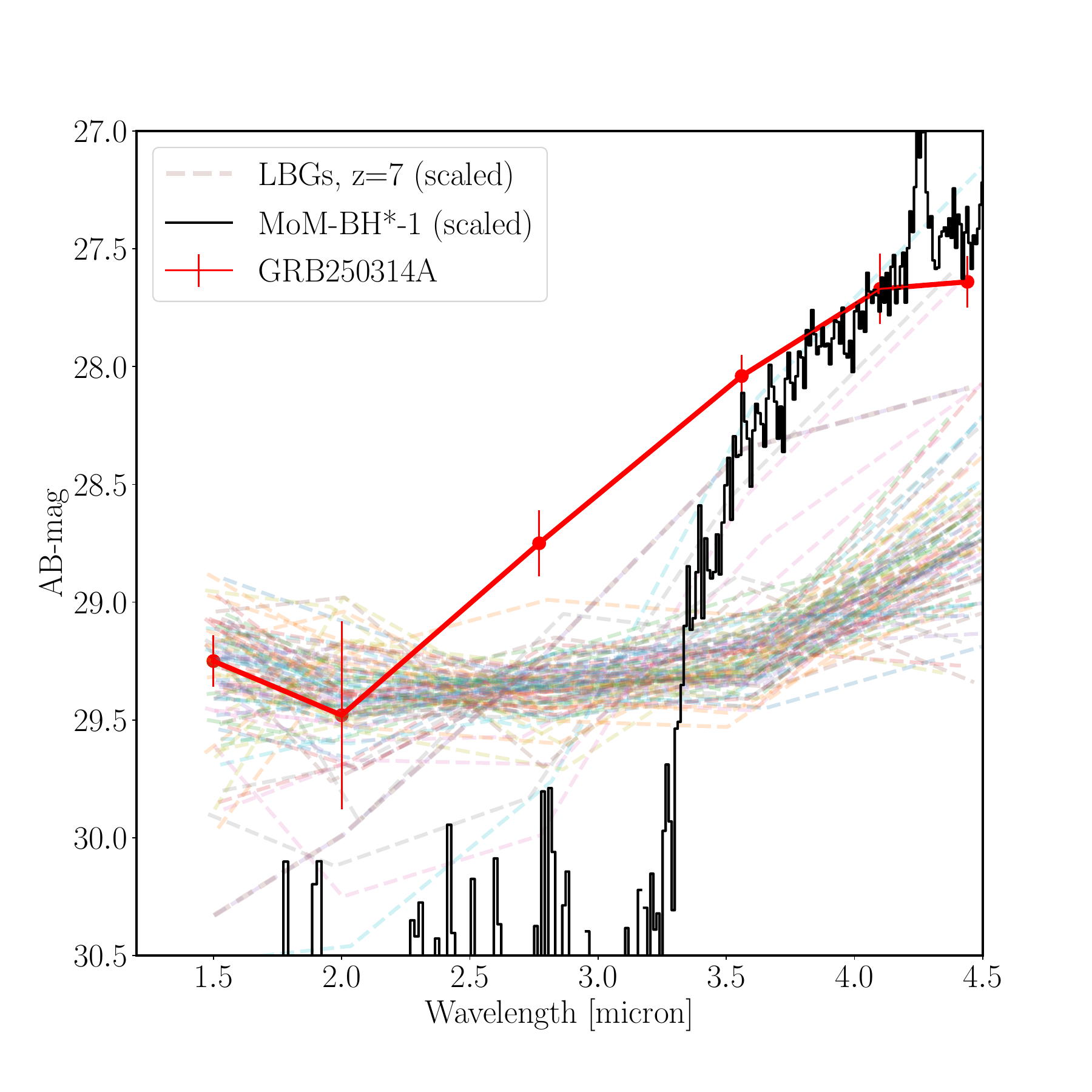}
\caption{The colours of 93 LBGs identified in the JADES, CEERS and PRIMER deep fields by \textit{JWST} \citep{Merlin2024} compared to the colours of GRB 250314A. Although a handful of LBGs do show apparent red colours these do not exhibit the same bluer shape in F150W-F200W, and indeed are dominated by Balmer break-like emission that increases predominantly beyond F277W. We also plot the NIRSPEC spectrum of the LRD MoM-BH1 \citep{naidu25}, which shows a very strong (likely non-stellar) Balmer break, and demonstrates that while very red objects can be found at $z \sim 7$, they do not explain the spectral shape seen in GRB 250314A.
}
\label{fig:lbg}
\end{figure}

In Figure~\ref{fig:lbg} we plot the SEDs of 93 Lyman break galaxies observed with NIRCAM at $z \sim 7$ \citep{Merlin2024}, scaled down by 2.2 magnitudes to match the blue end of the NIRCAM observations of GRB 250314A. As is apparent, these galaxies do not provide a reasonable match to the observations. Although a handful do show sufficiently red colours, the overall match to the spectral shape is poor. 

An alternative explanation would be to invoke a host similar to the population of little red dots \citep[LRDs,][]{matthee2024}, which are not thought to be explained by emission from a purely stellar population \citep{matthee2024,rusakov2025}. An LRD as a GRB host would be of substantial interest in its own right. LRDs comprise a minority of high-$z$ galaxies \citep[1--10\%,][]{Kokorev2024} and display blue continuum combined with a red excess beyond the Balmer break \citep{kocevski2025,killi2024}. This includes larger Balmer breaks at higher redshift than potentially seen here \citep[e.g.][]{Naidu2025}. However, all LRDs exhibit inflections at the Balmer limit \citep{setton2024}, at $z\simeq7.3$, the spectrum would be expected to begin rising in flux into the red wavelength region at $\sim3\,\mu$m, only at the edge of the F277W filter. In contrast the SN component flux rises very rapidly redward of 300\,nm, which is $\sim2.5\,\mu$m, providing substantial flux in the F277W band, comparable to the inferred host galaxy flux for a standard LBG SED. In other words, the rise of the red component observed in the F277W filter, is consistent with the SN interpretation, but not with a strong Balmer break as in LRDs. We conclude that known $z \sim 7$ galaxies, even with very strong Balmer breaks, are not consistent with the   light seen in the aftermath of GRB~250314A. 

However, it should also be considered, that GRB host selection differs from the selection function for galaxies found in wide-field surveys, and could sample different regions of the galaxy parameter space. 
Assuming that the observed SED is entirely dominated by host emission, we fit the available photometry with the SED fitting code CIGALE \citep{Boquien2019}. For the input parameters, we set the redshift at 7.3 \citep{2025GCN.39732....1M}. We adopt a delayed star formation history (SFH) with or without a recent burst. The main stellar population age is constrained to a maximum of 700~Myr, representing the limit of what is physically possible at $z=7.3$, with a maximum e-folding time of 200~Myr. The maximum age of the burst is defined at 100~Myr with a stellar mass fraction due to the burst $<$0.1. A \cite{Chabrier03} initial mass function with the stellar synthesis models of \cite{Bruzual03} is used, and models include nebular emission with standard parameters. The possible attenuation by dust is modeled using a modified version of the attenuation law of \cite{Calzetti00} assuming a maximum color excess of $E(B-V) = 0.2$. This choice is motivated by the relatively blue spectral slope of the GRB afterglow reported in \cite{cordier25}, which  indicates a low dust extinction along the line of sight and suggests a likely minimal dust attenuation within the host galaxy as also observed for star-forming galaxies at those redshifts \citep{Saxena24}.

The results demonstrate that it is possible to model the observed magnitudes of GRB\,250314A as purely arising from the host galaxy. 
However, the unsatisfactory aspects of this interpretation remain.
In particular, in order to explain the observed brightening between the F150W2 and F444W bands one must invoke a significant Balmer decrement, perhaps indicative of (i) old stars; or (as noted above) (ii) the class of recently discovered {\em JWST} Little Red Dots (LRDs). 

The output parameters are derived from the probability distribution functions (PDFs), using their mean values and associated standard deviations. The posterior PDFs favor a model without a recent burst and a population age of $562 \pm 71$ Myr, corresponding to an extremely early redshift of formation ($z \sim 21$). While not impossible, this is likely extremely rare and not entirely comfortable.
The resulting properties of the fit provide $\log(M_\ast / M_\odot) = 8.9 \pm 0.1$, and $\mathrm{SFR} = 0.4 \pm 0.2\,M_\odot\,\mathrm{yr}^{-1}$. This puts this galaxy well below the main sequence at these redshifts \citep{Popesso2023,Ciesla24} and yields a relatively low specific star formation rate ($\log(\mathrm{sSFR}) = -9.4 \pm 0.3\,\mathrm{yr}^{-1}$), indicating relatively low ongoing star formation relative to past star formation.
While such specific star formation rates are not unique in the $z \sim 7$ host galaxy population \citep{Ciesla24,roberts24}, it seems to be quite uncommon, and is not in keeping with what is expected and observed for GRB host galaxies \citep{perley16}.

Finally, as noted in the main text, in the host-only scenario, the similarity in the absolute magnitude of the source to SN~1998bw is entirely co-incidental. We therefore consider that, while possible in the absence of further observations to constrain variability, the most likely explanation of the available photometry is that we are detecting light from the associated supernova. 

\section*{Acknowledgements}
    This work is based on observations made with the NASA/ESA/CSA James Webb Space Telescope. The data were obtained from the Mikulski Archive for Space Telescopes at the Space Telescope Science Institute, which is operated by the Association of Universities for Research in Astronomy, Inc., under NASA contract NAS 5-03127 for JWST. These observations are associated with program 7296. Based on observations collected at the European Southern Observatory under ESO programme(s) 114.27PZ.
    BS, ELF, SDV and VB acknowledge the support of the French Agence Nationale de la Recherche (ANR), under grant ANR-23-CE31-0011 (project PEGaSUS).
    DBM, DW, and ASn are funded by the European Union (ERC, HEAVYMETAL, 101071865). The Cosmic Dawn Center (DAWN) is funded by the Danish National Research Foundation under grant DNRF140.
    ASa acknowledges support by a postdoctoral fellowship from the CNES. 
    BPG is supported by STFC grant No.
ST/Y002253/1 and Leverhulme Trust grant No. RPG-2024-117.
FEB acknowledges support from ANID-Chile BASAL CATA FB210003, FONDECYT Regular 1241005,
and Millennium Science Initiative, AIM23-0001.
NRT, NH acknowledge support from STFC grant ST/W000857/1. AMC and LC acknowledge support from the Irish Research Council Postgraduate Scholarship No. GOIPG/2022/1008.

\end{appendix}

\end{document}